\documentclass[fleqn,12pt,twoside]{article}
\usepackage{espcrc1}
\usepackage{graphicx}
\usepackage{wrapfig}
\newcommand{\slow}{$\sqrt{s_{NN}}$ = 130~GeV }
\newcommand{\shigh}{$\sqrt{s_{NN}}$ = 200~GeV }
\newcommand{\pt}{$p_T$ }

\title{Single leptons from heavy-flavor decays at RHIC}
\author{R. Averbeck\address{Department of Physics and Astronomy, 
State University of New York at Stony Brook, Stony Brook, NY 11794-3800, USA}
for the PHENIX Collaboration{\thanks{for the full PHENIX Collaboration author
list and acknowledgements, see Appendix "Collaborations" of this volume.}}}
       
\begin{document}

\maketitle

\begin{abstract}
Inclusive transverse momentum spectra of single electrons from Au+Au 
collisions have been measured at \slow and 200~GeV at midrapidity by the 
PHENIX experiment at RHIC.
After subtraction of background from photon conversions and light hadron 
decays, the spectra appear consistent with semileptonic decays of charmed
particles.
\end{abstract}

\section{Introduction}
\label{sec:introduction}
Particles carrying heavy flavor, {\it i.e.} charm or beauty quarks, are an
important probe of the hot and dense medium created in high energy 
heavy-ion collisions. 
Heavy-flavor production proceeds mainly via gluon-gluon fusion in the 
collision's earliest stage, thus being sensitive to the initial gluon density
\cite{App92,Mue92} and to nuclear effects, such as shadowing.
While propagating through the dense medium which could be in a deconfined
state, quarks can lose energy by gluon radiation \cite{Lin96,Dok01}. 
This might lead to a softening of final state particle spectra.
Furthermore, heavy-flavor measurements provide an important baseline
for the study of quarkonium suppression, which is a proposed signal of 
deconfinement \cite{Mat86}. 

The direct reconstruction of heavy-flavor decays,  
{\it e.g.} $D^0 \rightarrow K^- \pi^+$, is difficult in the high-multiplicity 
environment of a heavy-ion collision.
An alternative is to determine the contributions from semileptonic 
heavy-flavor decays, {\it e.g.} $D \rightarrow e K \nu$, to single lepton 
and lepton pair spectra.
PHENIX follows this approach in the analysis of single electrons,
$(e^+ + e^-)/2$, measured in Au+Au collisions at \slow \cite{Phe02} and 
200~GeV.

\section{Experiment and Analysis}
\label{sec:experiment}
The data used for this analysis were recorded by the PHENIX east-arm 
spectrometer ($\Delta\phi = 90^\circ$ in azimuth, $|\eta| < 0.35$ in 
pseudorapidity) which is described in detail elsewhere \cite{Ham02}.
The electron measurement employs a multitude of subdetectors \cite{Phe02}.
Beam-beam counters and zero-degree calorimeters provide the minimum bias
trigger, measure the vertex position, and are used for centrality selection.
Charged particle tracks are reconstructed with drift chambers and a layer of 
pad chambers.
Electron identification is performed with ring imaging Cerenkov detectors
and electromagnetic calorimeters.
The raw spectra are corrected for geometrical acceptance, reconstruction and 
particle-identification efficiency,
and for the multiplicity dependent efficiency loss due to detector occupancy.

\begin{wrapfigure}{l}{0.60\textwidth}
\begin{center}
\includegraphics[width=0.60\textwidth]{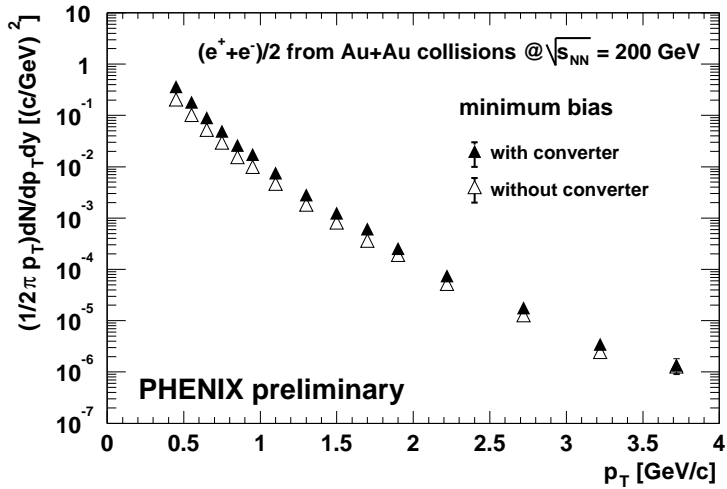}
\caption{Invariant \pt spectra of electrons from minimum bias Au+Au collisions
at $\sqrt{s_{NN}}$ = 200~GeV measured in PHENIX without and with the additional
photon converter installed.}
\label{fig:inclusive}
\end{center}
\end{wrapfigure}

The resulting invariant \pt distribution of electrons, measured from 
about 5.6$M$ minimum bias Au+Au collisions, is shown in 
Fig.~\ref{fig:inclusive}.
The main sources contributing to the electron spectra can be divided into two 
categories\footnote{The contribution from other sources, {\it e.g.} vector 
meson decays, is only marginal.}, {\it photonic} sources, {\it i.e.} 
conversion of photons in material in PHENIX and Dalitz decays of light mesons,
and {\it non-photonic} sources, {\it i.e.} semileptonic decays of heavy flavor.
To separate these from each other as described below, a data set of about
3.2$M$ minimum bias Au+Au collisions was used where a photon converter was
added to the standard PHENIX setup.
The converter, a thin brass tube with a radius of 29~cm, was installed in the 
center of PHENIX with the cylinder axis aligned with the beam line. 
The \pt distribution of electrons from the converter run is compared with the 
distribution measured without converter in Fig.~\ref{fig:inclusive}.
At low \pt the converter roughly doubles the electron yield.
Since the spectral shapes of electrons from Dalitz decays and from photon 
conversions are almost identical one would expect the two spectra in 
Fig.~\ref{fig:inclusive} to exhibit the same shape if all electrons 
were from photonic sources only.
The fact that the spectra approach each other at high \pt indicates the 
presence of a non-photonic source.

The non-photonic electron spectra are calculated in three steps.
First, the spectra of electrons originating from photon conversions in the 
converter itself are obtained by subtracting the spectra measured without 
converter from those measured with converter.
The result is corrected for the difference in material between the standard 
PHENIX setup and the setup with converter. 
Taking into account the relative contribution from Dalitz decays, the spectra 
of electrons from photonic sources in the standard setup are calculated from 
the pure conversion electron spectra in the second step.
Finally, the spectra of electrons from non-photonic sources are obtained by
subtracting the resulting electron spectra from photonic sources from the 
electron spectra measured in the standard setup.

\section{Results and Discussion}
\label{sec:results}
\begin{wrapfigure}{r}{0.6\textwidth}
\begin{center}
\includegraphics[width=0.6\textwidth]{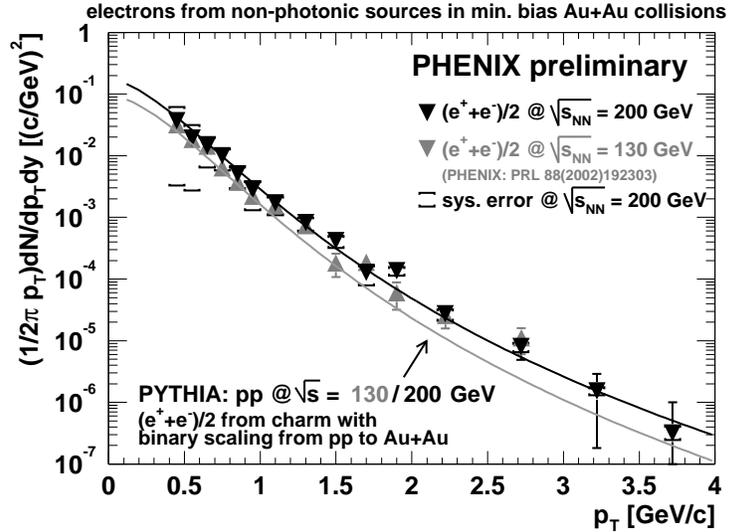}
\caption{Invariant \pt spectra of electrons from non-photonic sources in 
minimum bias Au+Au collisions at \shigh and $\sqrt{s_{NN}}$ = 130~GeV compared
with the expected contributions from semileptonic open charm decays.}
\label{fig:nonphotonic}
\end{center}
\end{wrapfigure}

The spectra of electrons from non-photonic sources in minimum bias Au+Au 
collisions at \shigh are shown in Fig.~\ref{fig:nonphotonic} together
with the corresponding spectra from 130~GeV \cite{Phe02}.
Note that the two data sets have been determined with two complementary 
methods with different systematic uncertainties, {\it i.e.} the converter 
method at 200~GeV and a cocktail subtraction method at 130~GeV.
The curves in Fig.~\ref{fig:nonphotonic} correspond to predictions  
of electron spectra from semileptonic charm decays as calculated with PYTHIA 
for $pp$ collisions scaled to Au+Au collisions using the number of binary 
collisions which are determined from a Glauber model calculation. 
The PYTHIA parameters have been tuned such that charm data from SPS and FNAL 
as well as single electron data from ISR are well described \cite{Phe02}.
The charm production cross section $\sigma_{c\bar{c}}$ in $pp$ collisions
from this PYTHIA calculation is 330~$\mu$b at $\sqrt{s}$ = 130~GeV and
650~$\mu$b at 200~GeV.
The electron yields and the spectral shapes are in reasonable agreement with 
the expectation from charm decays.

\begin{figure}[th]
\begin{center}
\begin{minipage}{\textwidth}
\includegraphics[width=0.5\textwidth]{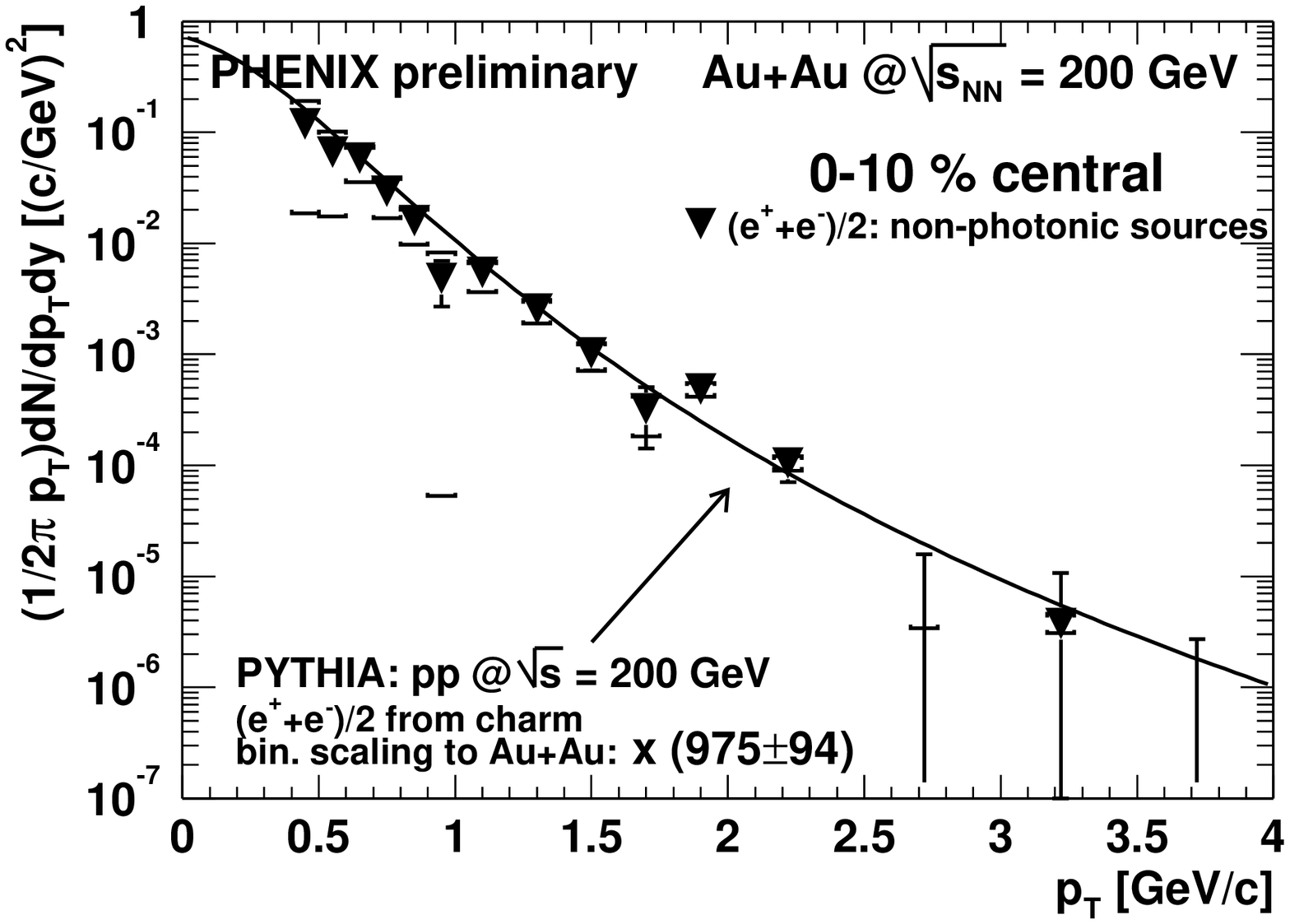}
\includegraphics[width=0.5\textwidth]{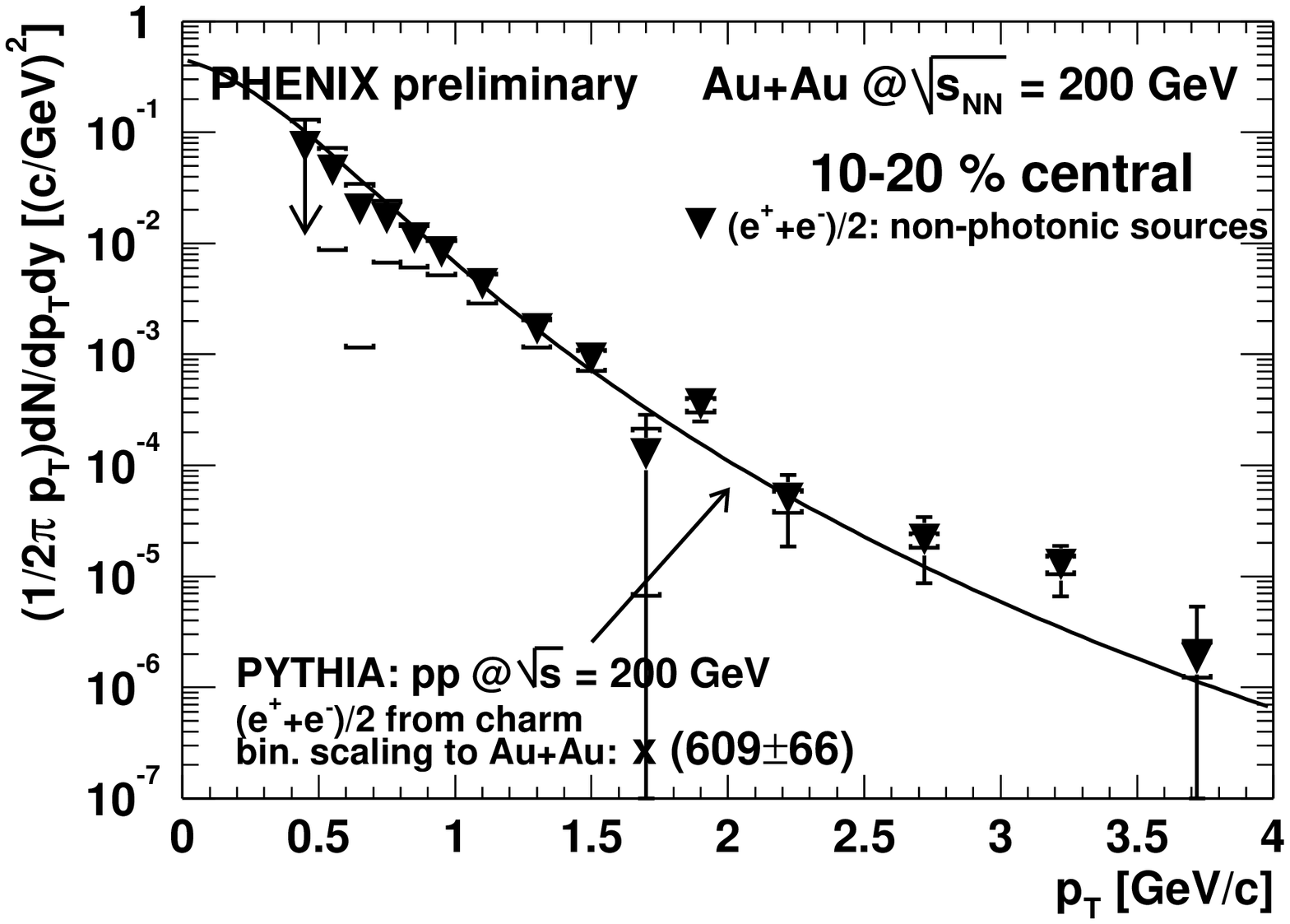}
\end{minipage}
\begin{minipage}{\textwidth}
\includegraphics[width=0.5\textwidth]{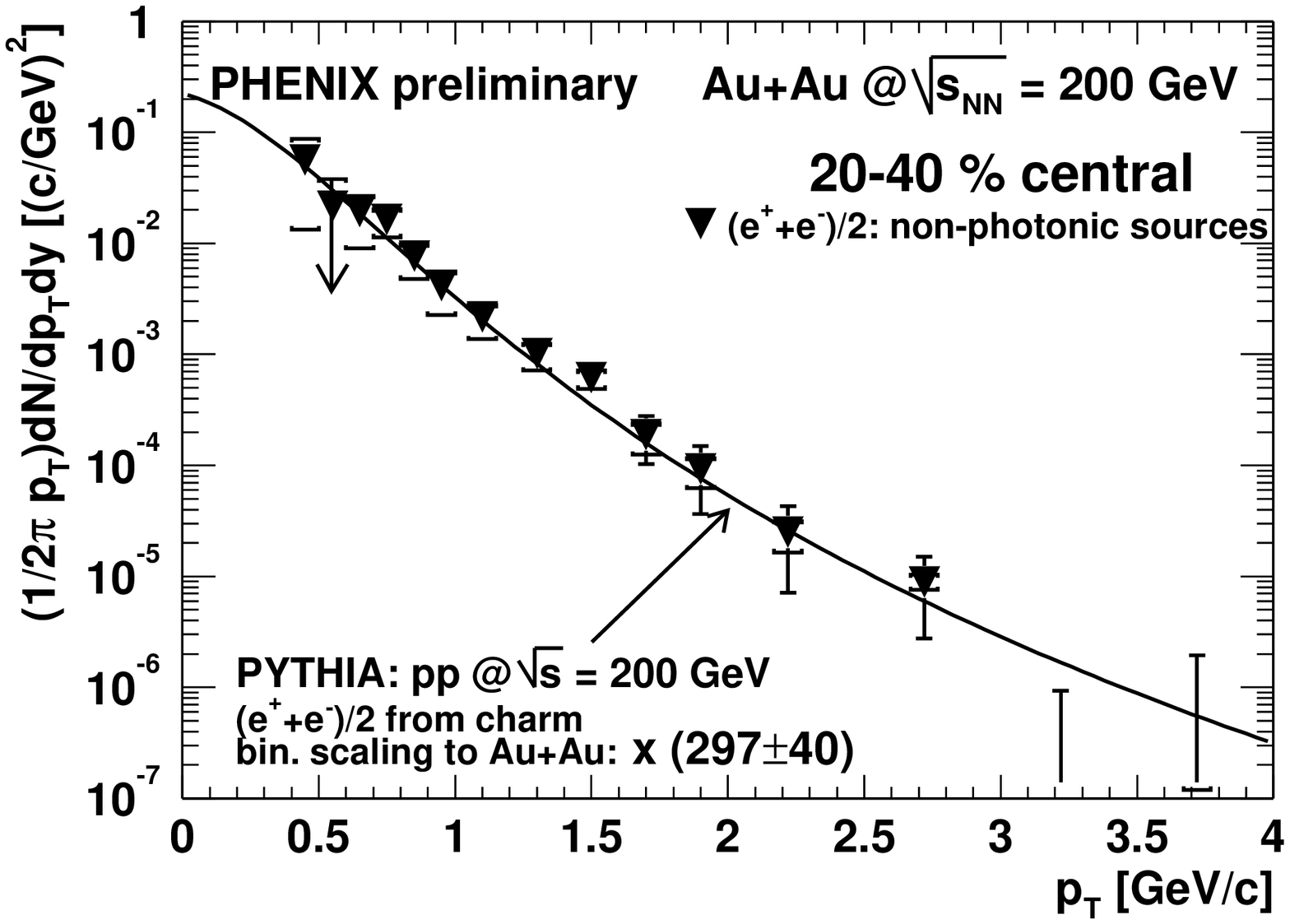}
\includegraphics[width=0.5\textwidth]{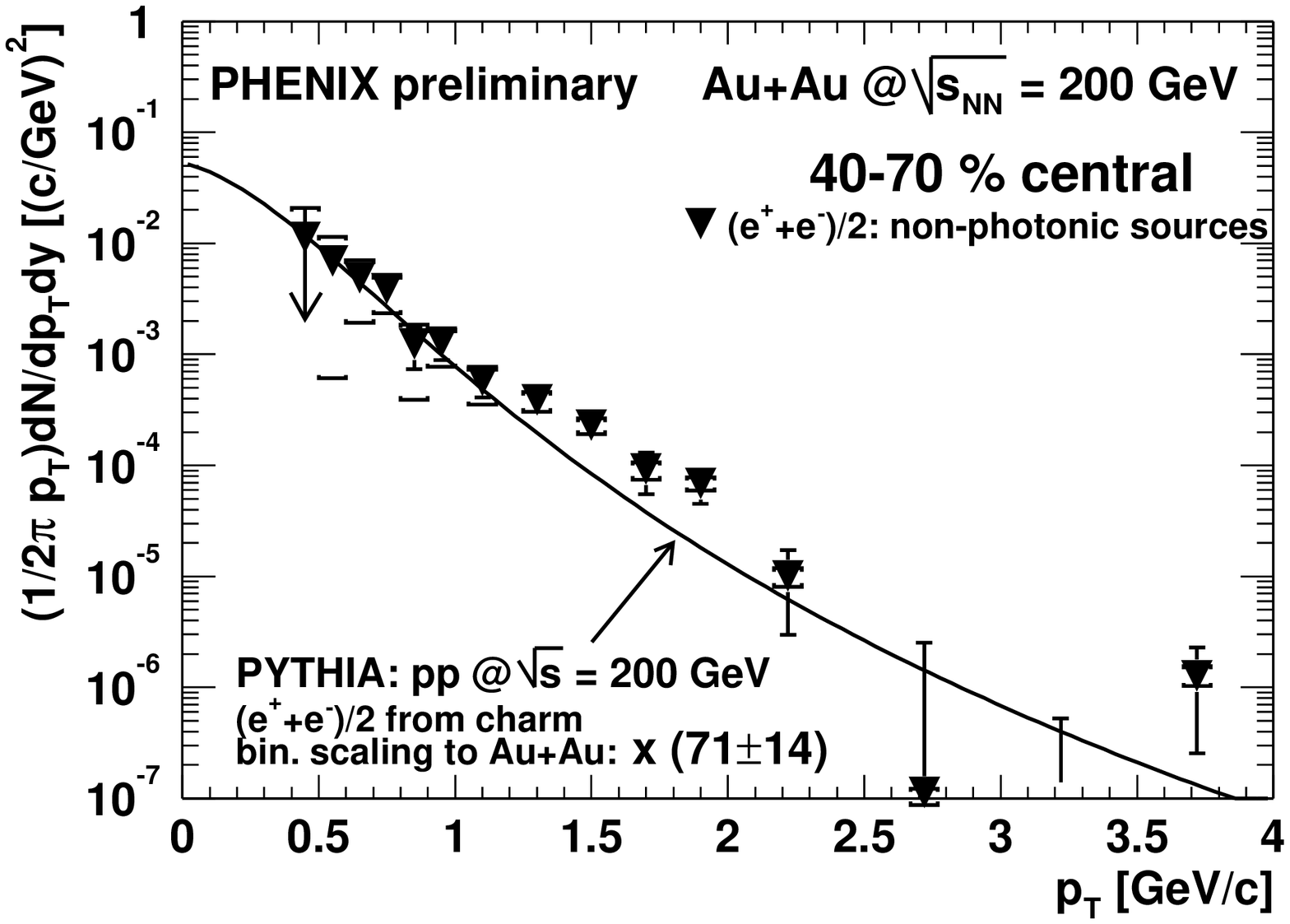}
\end{minipage}
\caption{Invariant \pt spectra of electrons from non-photonic sources in 
Au+Au collisions at \shigh compared with the expected contributions from 
semileptonic open charm decays for four different centrality selections.}
\label{fig:centrality}
\end{center}
\end{figure}

It is important to study the centrality dependence of these electron spectra
since any medium effect is expected to be more pronounced in more central 
collisions. 
The centrality dependence is addressed in Fig.~\ref{fig:centrality} which 
shows the comparison of the non-photonic electron \pt spectra measured in 
Au+Au collisions at \shigh with the expectation from charm decays for four 
different centrality selections, ranging from 0\% - 10\% (most central) to 
40\% - 70\% (most peripheral) of the total inelastic Au+Au cross section.
Again, the $pp$ PYTHIA calculation has been scaled to Au+Au using the number
of binary collisions which increases by more than a factor of ten going from 
the most peripheral to the most central collisions.
Although the uncertainties are quite large, the agreement between data and 
expectation is reasonable for all centrality selections.

\section{Summary and Outlook}
\label{sec:conclusion}
The spectra of electrons from non-photonic sources measured in Au+Au 
collisions at \slow and 200~GeV are consistent with the expectation from 
semileptonic charm decays as determined by $pp$ PYTHIA calculations scaled 
to Au+Au using the number of binary collisions.
At present, this is the only observable measured at RHIC obeying binary 
scaling within the experimental uncertainties.
At SPS, the NA50 Collaboration inferred an enhancement of the charm yield by 
a factor of about three in Pb+Pb collisions compared to binary collision 
scaled $pp$ measurement \cite{NA50}.
We do not observe such a large effect at RHIC.
In central Au+Au collisions at RHIC, a suppression of high \pt hadron yields 
by a factor of 3-4 was reported relative to binary scaling 
\cite{saskia,gerd}.
We do not observe such a large effect in the electron spectra from non-photonic
sources.
This could reflect a smaller energy loss of charm compared to light 
quarks, c.f. \cite{Dok01}.
For the \shigh data, both the statistical and the systematic uncertainties 
can still be reduced significantly, allowing for a more quantitative assessment
of heavy-flavor production at RHIC.
In particular, the analysis of the $pp$ data at $\sqrt{s}$ = 200~GeV will 
provide an actual  measurement as baseline for the search for eventual
medium effects instead of the PYTHIA prediction, thus reducing the model 
dependence of the interpretation significantly.

\end{document}